\documentclass[sigconf]{acmart}
\AtBeginDocument{%
  }

\copyrightyear{2025}
\acmYear{2025}
\setcopyright{rightsretained}
\acmConference[CHI EA '25]{Extended Abstracts of the CHI Conference on Human Factors in Computing Systems}{April 26-May 1, 2025}{Yokohama, Japan}
\acmBooktitle{Extended Abstracts of the CHI Conference on Human Factors in Computing Systems (CHI EA '25), April 26-May 1, 2025, Yokohama, Japan}\acmDOI{10.1145/3706599.3721183}
\acmISBN{979-8-4007-1395-8/2025/04}

\usepackage{siunitx}

\begin{document}

\title[Demo of picoRing \textit{mouse}]{Demo of picoRing \textit{mouse}: ultra-low-powered wireless mouse ring with ring-to-wristband coil-based impedance sensing}

\author{Yifan Li}
\affiliation{%
  \institution{The University of Tokyo}
  \city{Tokyo}
  \country{Japan}
  }
\email{yifan217@akg.t.u-tokyo.ac.jp}
\orcid{0009-0005-9261-6391}

\author{Masaaki Fukumoto}
\affiliation{%
  \institution{Microsoft Corporation}
  \city{Beijing}
  \country{China}
  }
\email{fukumoto@microsoft.com}
\orcid{1234-5678-9012}

\author{Mohamed Kari}
\affiliation{%
  \institution{Princeton University}
  \city{Princeton}
  \country{USA}
  }
\email{mk1809@cs.princeton.edu}
\orcid{0000-0003-4664-9983}

\author{Tomoyuki Yokota}
\affiliation{%
  \institution{The University of Tokyo}
  \city{Tokyo}
  \country{Japan}
  }
\email{yokota@ntech.t.u-tokyo.ac.jp}
\orcid{0000-0003-1546-8864}

\author{Takao Someya}
\affiliation{%
  \institution{The University of Tokyo}
  \city{Tokyo}
  \country{Japan}
  }
\email{someya@ee.t.u-tokyo.ac.jp}
\orcid{0000-0003-3051-1138}

\author{Yoshihiro Kawahara}
\affiliation{%
  \institution{The University of Tokyo}
  \city{Tokyo}
  \country{Japan}
  }
\email{kawahara@akg.t.u-tokyo.ac.jp}
\orcid{0000-0002-0310-2577}

\author{Ryo Takahashi}
\affiliation{%
  \institution{The University of Tokyo}
  \city{Tokyo}
  \country{Japan}
  }
\email{takahashi@akg.t.u-tokyo.ac.jp}
\orcid{0000-0001-5045-341X}

\renewcommand{\shortauthors}{Yifan Li, Ryo Takahashi, etc.}

\begin{abstract}
Wireless mouse rings offer subtle, reliable pointing interactions for wearable computing platforms.
However, the small battery below 27 mAh in the miniature rings restricts the ring's continuous lifespan to just 1-2 hours, because current low-powered wireless communication such as BLE is power-consuming for ring's continuous use. 
The ring's short lifespan persistently disrupts users' mouse use with the need for frequent charging.
This interactivity demonstrates picoRing \textit{mouse}, enabling a continuous ring-based mouse interaction with ultra-low-powered ring-to-wristband wireless communication. 
picoRing \textit{mouse} employs a coil-based impedance sensing named semi-passive inductive telemetry, allowing a wristband coil to capture a unique frequency response of a nearby ring coil via a sensitive inductive coupling between the coils.
The ring coil converts the corresponding user's mouse input into the unique frequency response via an 820 uW mouse-driven modulation module.
Therefore, continuous use of picoRing \textit{mouse} can potentially last over $92$ hours on a single charge of a $20$-mAh small battery. 
\end{abstract}



\received{20 February 2007}
\received[revised]{12 March 2009}
\received[accepted]{5 June 2009}

\maketitle

\section{INTRODUCTION}

\begin{figure*}[t]
  \centering
  \includegraphics[width=1.0\textwidth]{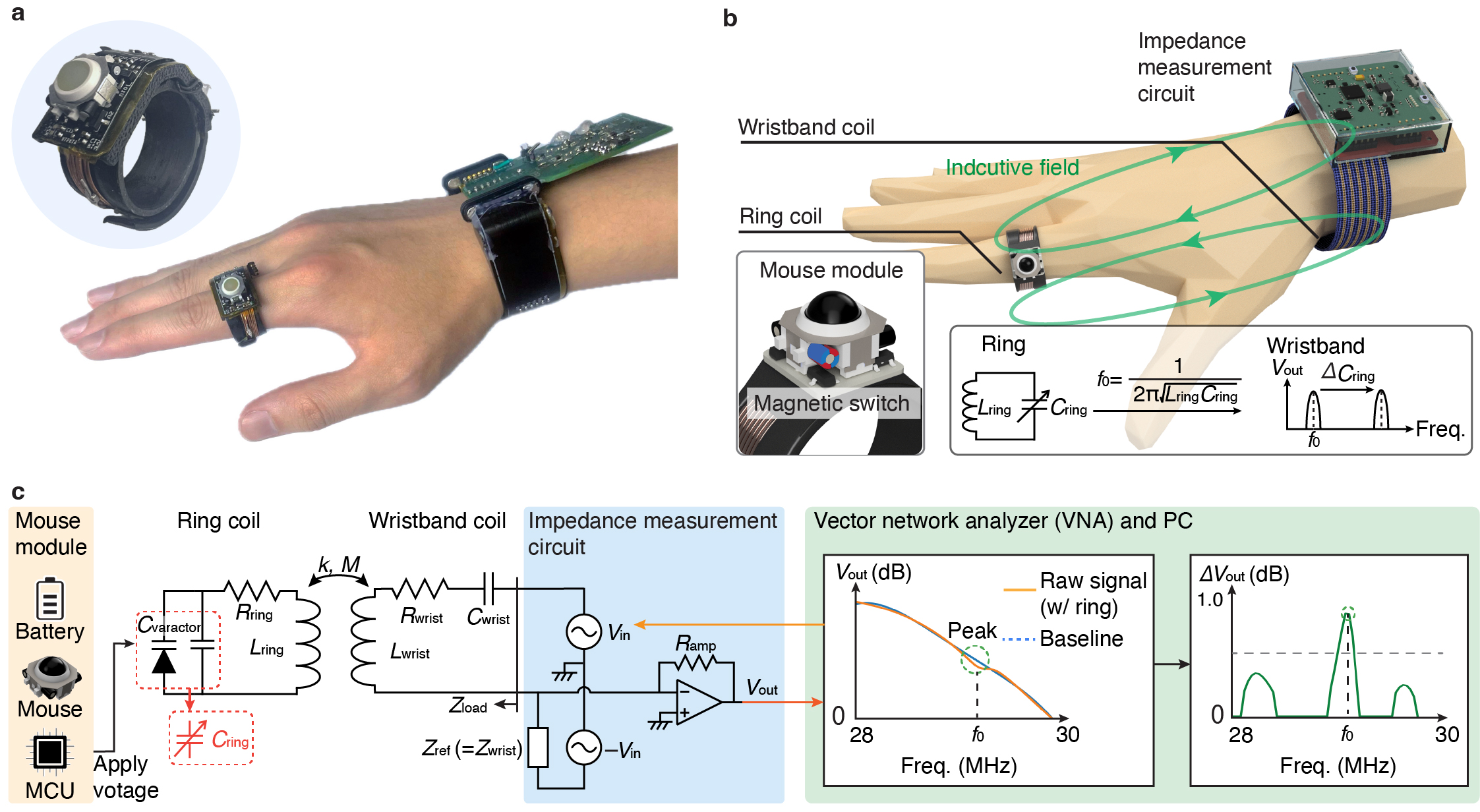}
  \caption{\textbf{Design overview of picoRing \textit{mouse}.} (a) Prototype photograph, (b) illustration, and (c) circuit diagram of picoRing \textit{mouse}.}
  \label{fig:overview}
  \Description{}
\end{figure*}

Mouse devices are basic tools for simple, fast pointing with computers.
Integrating mouse functionality into wearable form factors such as wristbands, glasses and rings (\textit{e.g.,} Apple Watch, Galaxy Ring) promises always-available essential interactions with wearable computing platforms~\cite{xu_ao-finger_2023, streli_touchinsight_2024, kienzle_electroring_2021}.
Especially, the wireless ring-formed input devices worn on the index finger offer reliable detection of even subtle, privacy-preserved thumb-to-index finger inputs such as pressing and scrolling, unlike the wristband and glasses~\cite{takahashi_picoring_2024,shen_mousering_2024,kim_iris_2024}.
However, the physical constraint of the tiny ring structure requires the equipment of tens of \si{mAh} small battery, challenging the continuous operation of power-consuming tens of \si{\mW}-class wireless communication modules such as Bluetooth Low Energy~(BLE).
For example, prior wireless rings support only $1-2$ hours of continuous wireless communication with BLE, limiting the ring's communication usage to intermittent data transfer~\cite{kim_iris_2024, shen_mousering_2024, fukumoto_body_1997}. 
Such an operation is suitable for periodic healthcare monitoring around the rings~(\textit{e.g.,} Oura Ring, Galaxy Ring), but not appropriate for ring-based input interface that requires both continuous interaction and real-time communication with other wearables~(\textit{e.g.,} smartwatch, HMD).

This Interactivity demonstrates picoRing \textit{mouse}, enabling an ultra-low-powered wireless mouse ring with a ring-to-wristband coil-based impedance sensing~(see \autoref{fig:overview}a).
picoRing \textit{mouse} is inspired by a sensitive coil-based impedance sensing named passive inductive telemetry~(PIT)~\cite{takahashi_picoring_2024, takahashi_telemetring_2020, takahashi_twin_2021}.
Unlike long-range electromagnetic communication such as BLE and RF backscatter, PIT constructs a short-range but ultra-low-powered inductive link between a pair of a ring coil and a wristband coil.
Since the ring coil can send its sensor information to the wristband coil by simply modifying the inductive field generated from the wristband coil, the ring coil does not need active signal transmission with power-hungry communication modules.
With the combination of PIT with the ring-based mouse module, picoRing \textit{mouse} enables the ultra-low-powered mouse ring that achieves \SI{800}{\uW} power consumption at most and weighs as little as \SI{5}{\gram}, possibly enabling continuous operation of over $92$ hours on a single charge of a \SI{20}{mAh} curved small battery.

\section{RELATED WORK}

Ring-formed mouse devices have long been investigated within HCI community.
The exploration of the previous mouse ring can be divided into two categories: i) ring-based sensing techniques such as computer vision, pressure, and inertial sensors to track the fine-grained finger movement~\cite{jiang_emerging_2022} and ii) low-powered wireless ring design using antenna- or coil-based low-power wireless communication such as RF backscatter and near-field communication~(NFC)~\cite{takahashi_picoring_2024, zhan_flexible_2024} to continuously operate the ring devices during daily life.
The former demonstrates high-fidelity microgesture recognition compared to the wristband- and glasses-based finger sensing~\cite{kienzle_electroring_2021}.
However, the latter is still challenging since \si{\uW}-class RF backscatter needs a large and unsuitable antenna in the ring~\cite{naderiparizi_towards_2018}, and \si{\mW}-class NFC offers a few centimeter-scale short communication that cannot construct ring-to-wristband link~\cite{lee_nfcstack_2022, zhu_robust_2024}.
picoRing, which is most similar to this paper, extends the coil-based communication distance by using coil-based sensitive impedance sensing for a ring-to-wristband low-powered wireless connectivity~\cite{takahashi_picoring_2024}.
While picoRing proposes a battery-free wireless ring, the input modality of the passive ring is limited to a few interfaces such as pressing~\cite{takahashi_picoring_2024} or tapping~\cite{takahashi_telemetring_2020}.
By contrast, picoRing \textit{mouse} can support multi-modal inputs in a single ring by integrating semi-passive communication architecture into the prior, as will be described in \S~\ref{sec:semi-PIT}.

\section{SYSTEM DESIGN}
\begin{figure*}[t!]
  \centering
  \includegraphics[width=1.0\textwidth]{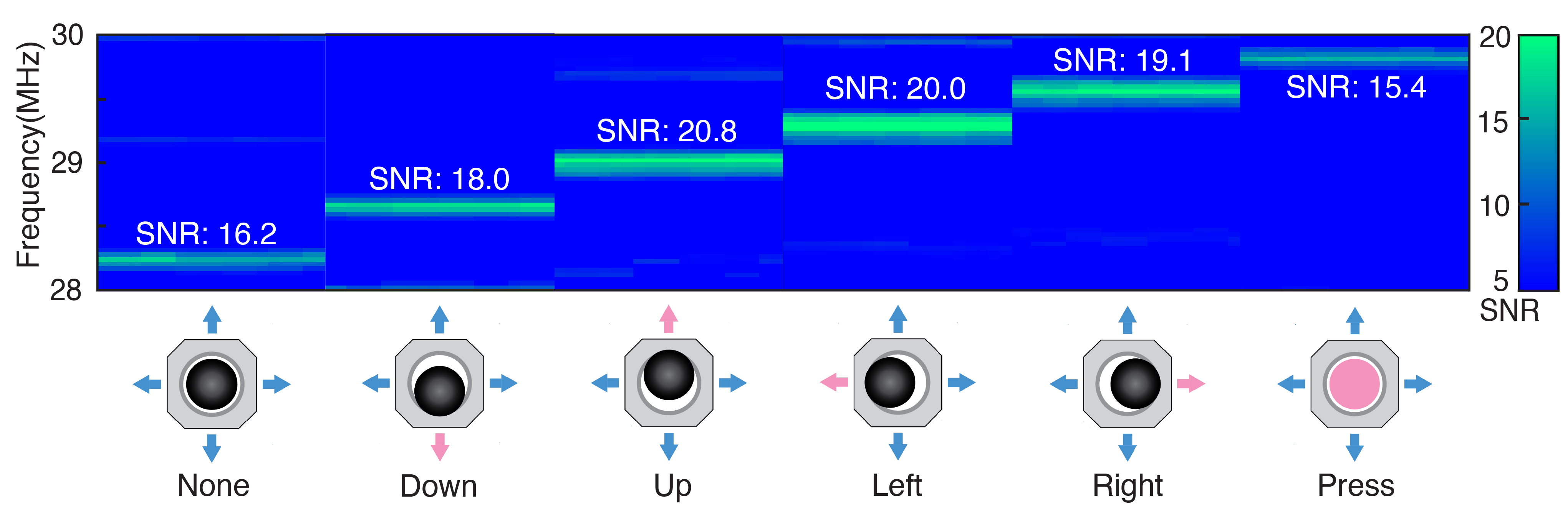}
  \caption{\textbf{Gesture recognition capability of picoRing \textit{mouse}.} We evaluate time-series and average SNR for scrolling and pressing interactions. SNR over $10$ indicates the high-fidelity recognition~\cite{takahashi_picoring_2024}.}
  \label{fig:evaluation}
  \Description{}
\end{figure*}

picoRing \textit{mouse} consists of two main components~(see ~\autoref{fig:overview}b): 1) a ring coil with a mouse module, which changes its frequency response based on the mouse inputs, and 2) a wristband coil that detects the peak in the frequency response corresponding to the ring's resonant frequency.
The working principle of picoRing \textit{mouse} is as follows:
First, the wristband coil generates a weak inductive field to couple with the nearby ring coil.
When the user scrolls the mouse module mounted on the ring, the MCU connected to the mouse module detects the scrolling direction via magnet switches, transforming the mouse direction into a unique frequency response~(\textit{i.e.,} the change in the ring's resonant frequency).
Through the inductive coupling, the wristband coil connected to a sensitive impedance measurement module obtains the frequency response.
Since the ring's frequency response is appeared as the frequency peak in the frequency characteristics of wristband's impedance, the wristband coil enables to recognize the mouse direction as a unique peak frequency shift.

\subsection{Semi-passive inductive telemetry}\label{sec:semi-PIT}

picoRing \textit{mouse} needs ultra-low-powered wireless communication for the long-term operation of the wireless ring mouse.
Among low-powered wireless communication, picoRing \textit{mouse} uses PIT similar to picoRing~\cite{takahashi_picoring_2024}.
PIT, which consists of a pair of a fully-passive ring coil and a wristband coil, allows the wristband to detect the passive response of the ring coil.
Since the ring-to-wristband inductive coupling is too weak, prior picoRing increases the PIT sensitivity by developing both a sensitive coil with distributed capacitors and a sensitive impedance measurement circuit named balanced bridge circuit.
The sensitive coil, which has a large turn number or large inductance at a high resonant frequency (\si{\MHz}) by inserting multiple chip capacitors into a single long coil, can increase the passive response from the ring coil.
By contrast, the bridge circuit, which uses a differential circuit structure, can be sensitive to the small impedance change from the wrist coil. 
For more detail, please refer to~\cite{takahashi_picoring_2024}~(see \autoref{fig:overview}c).
Note that the fully-passive features of the prior PIT requires the ring coil to modify the inductive field with a battery-free analog modulation circuit such as tactile switches.
In contrast, our semi-PIT approach, which modifies the inductive field via electrical switch, allows the ring coil to use digital modulation and digital signal processing, embedding the programmable, multi-modal interactions like our mouse-based scrolling and pressing into the inductive modulation.
Specifically, picoRing \textit{mouse} uses a simple frequency-shift keying based on a voltage-controlled varctor, which encodes the mouse inputs into a unique shift of ring's peak frequency~($f_0$) by the applied voltage magnitude.
Note that $f_0$ is a ring's resonant frequency determined by the ring's inductance~($L_{\rm ring}$) and capacitance~($L_{\rm ring}$) as follows: $f_0 = 1/(2\pi\sqrt{L_{\rm ring}C_{\rm ring}})$.

\begin{figure*}[t!]
  \centering
  \includegraphics[width=1.0\textwidth]{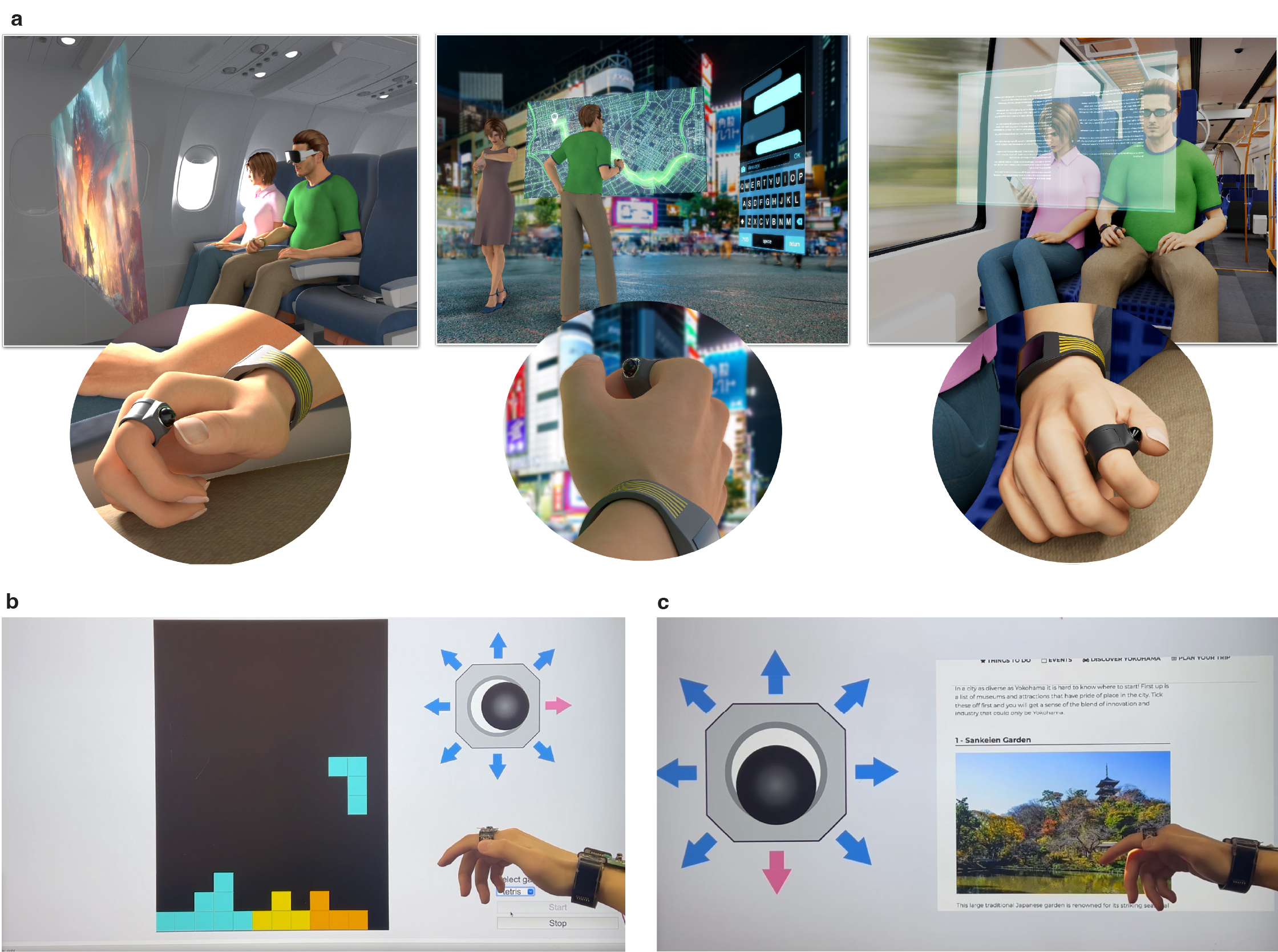}
  \caption{\textbf{Application examples of picoRing \textit{mouse}.} (a) Illustration of potential usage of picoRing \textit{mouse} during daily lives. (b) Photograph of two types of demonstration by picoRing \textit{mouse}. Wearable game controller for tetris playing and page scroll controller for web viewing.}
  \label{fig:demostration}
  \Description{}
\end{figure*}

\subsection{Ring}
\label{sec:ring}

The ring coil consists of a $8$-turned resonant coil with distributed capacitors and the mouse module equipped with a small trackball~(EVQWJN007, Panasonic) supporting up, down, right, left scrolling and pressing interactions. 
When users scroll or press the trackball in the mouse module, the trackball either converts the scroll rotation into rotation of cylindrical magnets with alternating polarity distribution or turns on the tactile button.
The magnetic switch (CT8132BL, Allegro MicroSystems) below each magnet sends the magnet rotation data to an ultra-low-power MCU (STM32L011F4U6, STM), and then, according to the input type, the MCU applies a different corresponding voltage to a varactor~(SMV1253, Skyworks) by a digital-potentiometer-based voltage divider circuit~(AD5160, Analog Devices). 
The varactor, which changes its capacitance ranging from \SI{27}{\pF} to \SI{69}{\pF} based on the applied voltage, changes the $C_{\rm ring}$ or $f_{0}$ by being connected in parallel to one of the distributed capacitors in the ring coil. 
In total, the ring coil provides five physical interface 1-D/2-D inputs: scroll up, down, left, right, scroll and press, and its resonance frequency for each interaction is tuned to \SI{28.0}{\MHz},\SI{28.4}{\MHz}, \SI{28.8}{\MHz}, \SI{29.2}{\MHz}, \SI{29.6}{\MHz}, respectively.
The inductance~($L_{\rm ring}$) and resistance~($R_{\rm ring}$) of the ring coil are \SI{2.6}{\uH} and \SI{3.5}{\ohm}, respectively.
Note that the ring coil is tilted with approximately $20^{\circ}$, mitigating the misalignment between the ring and wristband coils when the hand is grasped for thumb-to-index finger input.
With this tilted ring, the measured inductive coupling coefficient~($k$) increases from $0.0031$ to $0.0039$.

The ring coil has two types of operation modes: ACTIVE mode and STANDBY mode.
In the ACTIVE mode, the ring coil continuously streams data to the wristband coil.
By contrast, in the STANDBY mode, the timer in the MCU periodically awakens to check the receiving data from the connected magnet switch.
picoRing basically waits in the STANDBY mode and transits to ACTIVE mode when the user inputs to the mouse module. 
When there is no signal change in the magnet switch for \SI{5}{\s}, picoRing backs to the STANDBY mode.
The power consumption is approximately \SI{800}{\uW}~($=$\SI{1.8}{\V}$\times$\SI{430}{\uA}) and \SI{50}{\uW}~($=$\SI{1.8}{\V}$\times$\SI{28}{\uA}) in the ACTIVE and STANDBY modes, respectively.
The continuous ACTIVE operation time of picoRing \textit{mouse} is estimated to be approximately $92$ hours with the \SI{20}{mAh} curved Lipo battery~(UFX150732). 
Assuming the standard mouse operation time as about four hours per day, picoRing can potentially operate over a few weeks on a single charge.

\subsection{Wristband}
\label{sec:wrist}

The wristband coil consists of a $6$-turned flexible resonant coil mounted on a 3D-printed flexible wristband and the impedance measurement circuit connected to a mobile vector network analyzer~(VNA) (NanoVNA-H4, AURSINC). 
To recognize the ring's mouse state, the wristband coil detects the ring's frequency peak~($f_0$) by monitoring the frequency sweep signal ranging from \SI{27.8}{\MHz} to \SI{29}{\MHz} with an input power of \SI{0.2}{\mW}.
The coil is connected to the impedance measurement circuit via a magnetic connector. 
In total, the resonant frequency, inductance~($L_{\rm wrist}$) and resistance~($R_{\rm wrist}$) of the wristband coil connected with seventeen \SI{140}{\pF} distributed capacitors and a \SI{50}{\ohm} resistor in series are \SI{27}{\MHz}, \SI{4.0}{\uH}, and \SI{53}{\ohm}, respectively, and the measured power consumption of the circuit and VNA are \SI{0.46}{\W} and \SI{3.9}{\W}, respectively.

\section{PRELIMINARY EVALUATION}

To characterize the input accuracy of picoRing \textit{mouse}, this section evaluates the signal-to-noise ratio (SNR) of picoRing \textit{mouse} for user's scrolling and pressing interactions.
The SNR is measured similar to \cite{takahashi_picoring_2024}.
\autoref{fig:evaluation} shows the time-series SNR and the average SNR for the first author.
The result shows that the average SNR for each input is over \SI{10}{\dB}, indicating that picoRing \textit{mouse} can support high-fidelity recognition, according to \cite{takahashi_picoring_2024}.

\section{DEMONSTRATION}

picoRing \textit{mouse}, an ultra-low-powered wireless mouse ring, has the advantage of long-term operation with a single charge, unlike the prior wireless ring.
Therefore, picoRing \textit{mouse} can enable ubiquitous, seamless, finger inputs during prolonged daily life including VR/AR interactions in the airplane and train or outside~(see \autoref{fig:demostration}a).
This Interactivity demonstrates three application examples to highlight the effectiveness of picoRing \textit{mouse}~(see \autoref{fig:demostration}bc).
The first demonstration allows users to play and pause music by using a subtle pinch gesture on the ring's scroll ball. Because the picoRing \textit{mouse} can reliably detect subtle finger inputs over extended periods, users can frequently and naturally use the mouse ring without worrying about battery life.
The second demo shows wearable game controller experience~(see \autoref{fig:demostration}b).
Here, the user can control the rotation of various falling blocks in Tetris even for a day.
Owing to semi-PIT, picoRing \textit{mouse} can support complicated user interactions in addition to an ultra-low-powered operation. 
Lastly, the user scrolls the screen in AR/VR anywhere.
While the AR glasses and HMD support the scrolling interactions based on hand gestures, the gestures are relatively dynamic, causing fatigue in the continuous usage.
In contrast, the subtle finger input by picoRing \textit{mouse} allows the long-term privacy-preserving interactions.

\section{CONCLUSION}

This paper demonstrates picoRing \textit{mouse}, an ultra-low-power ring-based finger input device. 
By combining semiPIT-based ring-to-wristband low-powered wireless communication with the low-powered mouse module, picoRing \textit{mouse} achieves the \SI{800}{\uW} wireless ring mouse, supporting multi-modal, thumb-to-index, subtle finger inputs with long-term operation times of about $92$ hours.
As for the future works, picoRing \textit{mouse} could be used for spatial interaction in VR/AR by mathematically converting the user's subtle mouse input to extended 3D hand posture~\cite{kari_handycast_2023}.
Moreover, the power transmission of wireless charging-enabled clothing~\cite{takahashi_meander_2022, takahashi_cuttable_2018} can enable picoRing \textit{mouse} to work while being charged, allowing always wearing of a pair of the ring and wristband coils as fashion accessories.
We strongly believe picoRing \textit{mouse} could provide a new class of wearable computer mouse, seamlessly interconnecting itself with daily HCI devices.

\begin{acks}
This work was mainly supported by JST ACT-X JPMJAX21K9, JSPS KAKEN 22K21343, JST ASPIRE JPMJAP2401, and the Asahi Glass Foundation. 
\end{acks}

\bibliographystyle{ACM-Reference-Format}
\bibliography{references}


\begin{thebibliography}{17}


\ifx \showCODEN    \undefined \def \showCODEN     #1{\unskip}     \fi
\ifx \showDOI      \undefined \def \showDOI       #1{#1}\fi
\ifx \showISBNx    \undefined \def \showISBNx     #1{\unskip}     \fi
\ifx \showISBNxiii \undefined \def \showISBNxiii  #1{\unskip}     \fi
\ifx \showISSN     \undefined \def \showISSN      #1{\unskip}     \fi
\ifx \showLCCN     \undefined \def \showLCCN      #1{\unskip}     \fi
\ifx \shownote     \undefined \def \shownote      #1{#1}          \fi
\ifx \showarticletitle \undefined \def \showarticletitle #1{#1}   \fi
\ifx \showURL      \undefined \def \showURL       {\relax}        \fi
\providecommand\bibfield[2]{#2}
\providecommand\bibinfo[2]{#2}
\providecommand\natexlab[1]{#1}
\providecommand\showeprint[2][]{arXiv:#2}

\bibitem[Fukumoto and Tonomura(1997)]%
        {fukumoto_body_1997}
\bibfield{author}{\bibinfo{person}{Masaaki Fukumoto} {and} \bibinfo{person}{Yoshinobu Tonomura}.} \bibinfo{year}{1997}\natexlab{}.
\newblock \showarticletitle{“{Body} coupled {FingerRing}”: wireless wearable keyboard}. In \bibinfo{booktitle}{\emph{Proceedings of the {ACM} {SIGCHI} {Conference} on {Human} factors in computing systems}}. \bibinfo{publisher}{ACM}, \bibinfo{address}{Atlanta Georgia USA}, \bibinfo{pages}{147--154}.
\newblock
\showISBNx{978-0-89791-802-2}
\urldef\tempurl%
\url{https://doi.org/10.1145/258549.258636}
\showDOI{\tempurl}


\bibitem[Jiang et~al\mbox{.}(2022)]%
        {jiang_emerging_2022}
\bibfield{author}{\bibinfo{person}{Shuo Jiang}, \bibinfo{person}{Peiqi Kang}, \bibinfo{person}{Xinyu Song}, \bibinfo{person}{Benny~P.L. Lo}, {and} \bibinfo{person}{Peter~B. Shull}.} \bibinfo{year}{2022}\natexlab{}.
\newblock \showarticletitle{Emerging {Wearable} {Interfaces} and {Algorithms} for {Hand} {Gesture} {Recognition}: {A} {Survey}}.
\newblock \bibinfo{journal}{\emph{IEEE Reviews in Biomedical Engineering}}  \bibinfo{volume}{15} (\bibinfo{year}{2022}), \bibinfo{pages}{85--102}.
\newblock
\showISSN{1941-1189}
\urldef\tempurl%
\url{https://doi.org/10.1109/RBME.2021.3078190}
\showDOI{\tempurl}
\newblock
\shownote{Conference Name: IEEE Reviews in Biomedical Engineering}.


\bibitem[Kari and Holz(2023)]%
        {kari_handycast_2023}
\bibfield{author}{\bibinfo{person}{Mohamed Kari} {and} \bibinfo{person}{Christian Holz}.} \bibinfo{year}{2023}\natexlab{}.
\newblock \showarticletitle{{HandyCast}: {Phone}-based {Bimanual} {Input} for {Virtual} {Reality} in {Mobile} and {Space}-{Constrained} {Settings} via {Pose}-and-{Touch} {Transfer}}. In \bibinfo{booktitle}{\emph{Proceedings of the 2023 {CHI} {Conference} on {Human} {Factors} in {Computing} {Systems}}}. \bibinfo{publisher}{ACM}, \bibinfo{address}{Hamburg Germany}, \bibinfo{pages}{1--15}.
\newblock
\showISBNx{978-1-4503-9421-5}
\urldef\tempurl%
\url{https://doi.org/10.1145/3544548.3580677}
\showDOI{\tempurl}


\bibitem[Kienzle et~al\mbox{.}(2021)]%
        {kienzle_electroring_2021}
\bibfield{author}{\bibinfo{person}{Wolf Kienzle}, \bibinfo{person}{Eric Whitmire}, \bibinfo{person}{Chris Rittaler}, {and} \bibinfo{person}{Hrvoje Benko}.} \bibinfo{year}{2021}\natexlab{}.
\newblock \showarticletitle{{ElectroRing}: {Subtle} {Pinch} and {Touch} {Detection} with a {Ring}}. In \bibinfo{booktitle}{\emph{Proceedings of the 2021 {CHI} {Conference} on {Human} {Factors} in {Computing} {Systems}}}. \bibinfo{publisher}{ACM}, \bibinfo{address}{Yokohama Japan}, \bibinfo{pages}{1--12}.
\newblock
\showISBNx{978-1-4503-8096-6}
\urldef\tempurl%
\url{https://doi.org/10.1145/3411764.3445094}
\showDOI{\tempurl}


\bibitem[Kim et~al\mbox{.}(2024)]%
        {kim_iris_2024}
\bibfield{author}{\bibinfo{person}{Maruchi Kim}, \bibinfo{person}{Antonio Glenn}, \bibinfo{person}{Bandhav Veluri}, \bibinfo{person}{Yunseo Lee}, \bibinfo{person}{Eyoel Gebre}, \bibinfo{person}{Aditya Bagaria}, \bibinfo{person}{Shwetak Patel}, {and} \bibinfo{person}{Shyamnath Gollakota}.} \bibinfo{year}{2024}\natexlab{}.
\newblock \showarticletitle{{IRIS}: {Wireless} ring for vision-based smart home interaction}. In \bibinfo{booktitle}{\emph{Proceedings of the 37th {Annual} {ACM} {Symposium} on {User} {Interface} {Software} and {Technology}}}. \bibinfo{publisher}{ACM}, \bibinfo{address}{Pittsburgh PA USA}, \bibinfo{pages}{1--16}.
\newblock
\showISBNx{9798400706288}
\urldef\tempurl%
\url{https://doi.org/10.1145/3654777.3676327}
\showDOI{\tempurl}


\bibitem[Lee et~al\mbox{.}(2022)]%
        {lee_nfcstack_2022}
\bibfield{author}{\bibinfo{person}{Chi-Jung Lee}, \bibinfo{person}{Rong-Hao Liang}, \bibinfo{person}{Ling-Chien Yang}, \bibinfo{person}{Chi-Huan Chiang}, \bibinfo{person}{Te-Yen Wu}, {and} \bibinfo{person}{Bing-Yu Chen}.} \bibinfo{year}{2022}\natexlab{}.
\newblock \showarticletitle{{NFCStack}: {Identifiable} {Physical} {Building} {Blocks} that {Support} {Concurrent} {Construction} and {Frictionless} {Interaction}}. In \bibinfo{booktitle}{\emph{Proceedings of the 35th {Annual} {ACM} {Symposium} on {User} {Interface} {Software} and {Technology}}}. \bibinfo{publisher}{ACM}, \bibinfo{address}{Bend OR USA}, \bibinfo{pages}{1--12}.
\newblock
\showISBNx{978-1-4503-9320-1}
\urldef\tempurl%
\url{https://doi.org/10.1145/3526113.3545658}
\showDOI{\tempurl}


\bibitem[Naderiparizi et~al\mbox{.}(2018)]%
        {naderiparizi_towards_2018}
\bibfield{author}{\bibinfo{person}{Saman Naderiparizi}, \bibinfo{person}{Mehrdad Hessar}, \bibinfo{person}{Vamsi Talla}, \bibinfo{person}{Shyamnath Gollakota}, {and} \bibinfo{person}{Joshua~R. Smith}.} \bibinfo{year}{2018}\natexlab{}.
\newblock \showarticletitle{Towards \{{Battery}-{Free}\} \{{HD}\} {Video} {Streaming}}. \bibinfo{pages}{233--247}.
\newblock
\showISBNx{978-1-939133-01-4}
\urldef\tempurl%
\url{https://www.usenix.org/conference/nsdi18/presentation/naderiparizi}
\showURL{%
\tempurl}


\bibitem[Shen et~al\mbox{.}(2024)]%
        {shen_mousering_2024}
\bibfield{author}{\bibinfo{person}{Xiyuan Shen}, \bibinfo{person}{Chun Yu}, \bibinfo{person}{Xutong Wang}, \bibinfo{person}{Chen Liang}, \bibinfo{person}{Haozhan Chen}, {and} \bibinfo{person}{Yuanchun Shi}.} \bibinfo{year}{2024}\natexlab{}.
\newblock \showarticletitle{{MouseRing}: {Always}-available {Touchpad} {Interaction} with {IMU} {Rings}}. In \bibinfo{booktitle}{\emph{Proceedings of the {CHI} {Conference} on {Human} {Factors} in {Computing} {Systems}}}. \bibinfo{publisher}{ACM}, \bibinfo{address}{Honolulu HI USA}, \bibinfo{pages}{1--19}.
\newblock
\showISBNx{9798400703300}
\urldef\tempurl%
\url{https://doi.org/10.1145/3613904.3642225}
\showDOI{\tempurl}


\bibitem[Streli et~al\mbox{.}(2024)]%
        {streli_touchinsight_2024}
\bibfield{author}{\bibinfo{person}{Paul Streli}, \bibinfo{person}{Mark Richardson}, \bibinfo{person}{Fadi Botros}, \bibinfo{person}{Shugao Ma}, \bibinfo{person}{Robert Wang}, {and} \bibinfo{person}{Christian Holz}.} \bibinfo{year}{2024}\natexlab{}.
\newblock \showarticletitle{{TouchInsight}: {Uncertainty}-aware {Rapid} {Touch} and {Text} {Input} for {Mixed} {Reality} from {Egocentric} {Vision}}. In \bibinfo{booktitle}{\emph{Proceedings of the 37th {Annual} {ACM} {Symposium} on {User} {Interface} {Software} and {Technology}}}. \bibinfo{publisher}{ACM}, \bibinfo{address}{Pittsburgh PA USA}, \bibinfo{pages}{1--16}.
\newblock
\showISBNx{9798400706288}
\urldef\tempurl%
\url{https://doi.org/10.1145/3654777.3676330}
\showDOI{\tempurl}


\bibitem[Takahashi et~al\mbox{.}(2020)]%
        {takahashi_telemetring_2020}
\bibfield{author}{\bibinfo{person}{Ryo Takahashi}, \bibinfo{person}{Masaaki Fukumoto}, \bibinfo{person}{Changyo Han}, \bibinfo{person}{Takuya Sasatani}, \bibinfo{person}{Yoshiaki Narusue}, {and} \bibinfo{person}{Yoshihiro Kawahara}.} \bibinfo{year}{2020}\natexlab{}.
\newblock \showarticletitle{{TelemetRing}: {A} {Batteryless} and {Wireless} {Ring}-shaped {Keyboard} using {Passive} {Inductive} {Telemetry}}. In \bibinfo{booktitle}{\emph{Proceedings of the 33rd {Annual} {ACM} {Symposium} on {User} {Interface} {Software} and {Technology}}}. \bibinfo{publisher}{ACM}, \bibinfo{address}{Virtual Event USA}, \bibinfo{pages}{1161--1168}.
\newblock
\showISBNx{978-1-4503-7514-6}
\urldef\tempurl%
\url{https://doi.org/10.1145/3379337.3415873}
\showDOI{\tempurl}


\bibitem[Takahashi et~al\mbox{.}(2018)]%
        {takahashi_cuttable_2018}
\bibfield{author}{\bibinfo{person}{Ryo Takahashi}, \bibinfo{person}{Takuya Sasatani}, \bibinfo{person}{Fuminori Okuya}, \bibinfo{person}{Yoshiaki Narusue}, {and} \bibinfo{person}{Yoshihiro Kawahara}.} \bibinfo{year}{2018}\natexlab{}.
\newblock \showarticletitle{A {Cuttable} {Wireless} {Power} {Transfer} {Sheet}}.
\newblock \bibinfo{journal}{\emph{Proceedings of the ACM on Interactive, Mobile, Wearable and Ubiquitous Technologies}} \bibinfo{volume}{2}, \bibinfo{number}{4} (\bibinfo{date}{Dec.} \bibinfo{year}{2018}), \bibinfo{pages}{1--25}.
\newblock
\showISSN{2474-9567}
\urldef\tempurl%
\url{https://doi.org/10.1145/3287068}
\showDOI{\tempurl}


\bibitem[Takahashi et~al\mbox{.}(2024)]%
        {takahashi_picoring_2024}
\bibfield{author}{\bibinfo{person}{Ryo Takahashi}, \bibinfo{person}{Eric Whitmire}, \bibinfo{person}{Roger Boldu}, \bibinfo{person}{Shiu Ng}, \bibinfo{person}{Wolf Kienzle}, {and} \bibinfo{person}{Hrvoje Benko}.} \bibinfo{year}{2024}\natexlab{}.
\newblock \showarticletitle{{picoRing}: battery-free rings for subtle thumb-to-index input}. In \bibinfo{booktitle}{\emph{Proceedings of the 37th {Annual} {ACM} {Symposium} on {User} {Interface} {Software} and {Technology}}}. \bibinfo{publisher}{ACM}, \bibinfo{address}{Pittsburgh PA USA}, \bibinfo{pages}{1--11}.
\newblock
\showISBNx{9798400706288}
\urldef\tempurl%
\url{https://doi.org/10.1145/3654777.3676365}
\showDOI{\tempurl}


\bibitem[Takahashi et~al\mbox{.}(2021)]%
        {takahashi_twin_2021}
\bibfield{author}{\bibinfo{person}{Ryo Takahashi}, \bibinfo{person}{Wakako Yukita}, \bibinfo{person}{Takuya Sasatani}, \bibinfo{person}{Tomoyuki Yokota}, \bibinfo{person}{Takao Someya}, {and} \bibinfo{person}{Yoshihiro Kawahara}.} \bibinfo{year}{2021}\natexlab{}.
\newblock \showarticletitle{Twin {Meander} {Coil}: {Sensitive} {Readout} of {Battery}-free {On}-body {Wireless} {Sensors} {Using} {Body}-scale {Meander} {Coils}}.
\newblock \bibinfo{journal}{\emph{Proceedings of the ACM on Interactive, Mobile, Wearable and Ubiquitous Technologies}} \bibinfo{volume}{5}, \bibinfo{number}{4} (\bibinfo{date}{Dec.} \bibinfo{year}{2021}), \bibinfo{pages}{1--21}.
\newblock
\showISSN{2474-9567}
\urldef\tempurl%
\url{https://doi.org/10.1145/3494996}
\showDOI{\tempurl}


\bibitem[Takahashi et~al\mbox{.}(2022)]%
        {takahashi_meander_2022}
\bibfield{author}{\bibinfo{person}{Ryo Takahashi}, \bibinfo{person}{Wakako Yukita}, \bibinfo{person}{Tomoyuki Yokota}, \bibinfo{person}{Takao Someya}, {and} \bibinfo{person}{Yoshihiro Kawahara}.} \bibinfo{year}{2022}\natexlab{}.
\newblock \showarticletitle{Meander {Coil}++: {A} {Body}-scale {Wireless} {Power} {Transmission} {Using} {Safe}-to-body and {Energy}-efficient {Transmitter} {Coil}}. In \bibinfo{booktitle}{\emph{{CHI} {Conference} on {Human} {Factors} in {Computing} {Systems}}}. \bibinfo{publisher}{ACM}, \bibinfo{address}{New Orleans LA USA}, \bibinfo{pages}{1--12}.
\newblock
\showISBNx{978-1-4503-9157-3}
\urldef\tempurl%
\url{https://doi.org/10.1145/3491102.3502119}
\showDOI{\tempurl}


\bibitem[Xu et~al\mbox{.}(2023)]%
        {xu_ao-finger_2023}
\bibfield{author}{\bibinfo{person}{Chenhan Xu}, \bibinfo{person}{Bing Zhou}, \bibinfo{person}{Gurunandan Krishnan}, {and} \bibinfo{person}{Shree Nayar}.} \bibinfo{year}{2023}\natexlab{}.
\newblock \showarticletitle{{AO}-{Finger}: {Hands}-free {Fine}-grained {Finger} {Gesture} {Recognition} via {Acoustic}-{Optic} {Sensor} {Fusing}}. In \bibinfo{booktitle}{\emph{Proceedings of the 2023 {CHI} {Conference} on {Human} {Factors} in {Computing} {Systems}}}. \bibinfo{publisher}{ACM}, \bibinfo{address}{Hamburg Germany}, \bibinfo{pages}{1--14}.
\newblock
\showISBNx{978-1-4503-9421-5}
\urldef\tempurl%
\url{https://doi.org/10.1145/3544548.3581264}
\showDOI{\tempurl}


\bibitem[Zhan et~al\mbox{.}(2024)]%
        {zhan_flexible_2024}
\bibfield{author}{\bibinfo{person}{Jun-Lin Zhan}, \bibinfo{person}{Wei-Bing Lu}, \bibinfo{person}{Cong Ding}, \bibinfo{person}{Zhen Sun}, \bibinfo{person}{Bu-Yun Yu}, \bibinfo{person}{Lu Ju}, \bibinfo{person}{Xin-Hua Liang}, \bibinfo{person}{Zhao-Min Chen}, \bibinfo{person}{Hao Chen}, \bibinfo{person}{Yong-Hao Jia}, \bibinfo{person}{Zhen-Guo Liu}, {and} \bibinfo{person}{Tie-Jun Cui}.} \bibinfo{year}{2024}\natexlab{}.
\newblock \showarticletitle{Flexible and wearable battery-free backscatter wireless communication system for colour imaging}.
\newblock \bibinfo{journal}{\emph{npj Flexible Electronics}} \bibinfo{volume}{8}, \bibinfo{number}{1} (\bibinfo{date}{March} \bibinfo{year}{2024}), \bibinfo{pages}{1--9}.
\newblock
\showISSN{2397-4621}
\urldef\tempurl%
\url{https://doi.org/10.1038/s41528-024-00304-4}
\showDOI{\tempurl}
\newblock
\shownote{Publisher: Nature Publishing Group}.


\bibitem[Zhu et~al\mbox{.}(2024)]%
        {zhu_robust_2024}
\bibfield{author}{\bibinfo{person}{Xia Zhu}, \bibinfo{person}{Ke Wu}, \bibinfo{person}{Xiaohang Xie}, \bibinfo{person}{Stephan~W. Anderson}, {and} \bibinfo{person}{Xin Zhang}.} \bibinfo{year}{2024}\natexlab{}.
\newblock \showarticletitle{A robust near-field body area network based on coaxially-shielded textile metamaterial}.
\newblock \bibinfo{journal}{\emph{Nature Communications}} \bibinfo{volume}{15}, \bibinfo{number}{1} (\bibinfo{date}{Aug.} \bibinfo{year}{2024}), \bibinfo{pages}{6589}.
\newblock
\showISSN{2041-1723}
\urldef\tempurl%
\url{https://doi.org/10.1038/s41467-024-51061-x}
\showDOI{\tempurl}
\newblock
\shownote{Publisher: Nature Publishing Group}.


\end{thebibliography}

\appendix

\end{document}